\documentclass[sigconf]{acmart}
\usepackage{makecell}
\usepackage{dsfont}
\usepackage{multirow}
\usepackage{diagbox}
\AtBeginDocument{%
  }

\setcopyright{acmlicensed}
\copyrightyear{2026}
\acmYear{2026}
\setcopyright{cc}
\setcctype{by}
\acmConference[SIGIR '26]{Proceedings of the 49th International ACM SIGIR Conference on Research and Development in Information Retrieval}{July 20--24, 2026}{Melbourne, VIC, Australia}
\acmBooktitle{Proceedings of the 49th International ACM SIGIR Conference on Research and Development in Information Retrieval (SIGIR '26), July 20--24, 2026, Melbourne, VIC, Australia}
\acmDOI{10.1145/3805712.3808434}
\acmISBN{979-8-4007-2599-9/2026/07}

\begin{document}

\title{MMRM: A Multiplex Multimodal Representation Model for Product Ranking in E-commerce Search}

\author{Zhen-Lin Chen}
\orcid{0000-0003-0838-8348}
\affiliation{%
  \institution{JD.COM}
  \city{Beijing}
  \country{China}}
\email{chenzhenlin6@jd.com}

\author{Maosen Sheng}
\orcid{0009-0000-8910-3388}
\affiliation{%
	\institution{JD.COM}
	\city{Beijing}
	\country{China}}
\email{shengmaosen.1@jd.com}

\author{Peng Lin}
\orcid{0009-0007-3844-460X}
\affiliation{%
	\institution{JD.COM}
	\city{Beijing}
	\country{China}}
\email{linpeng47@jd.com}

\author{Jianmin Chen}
\orcid{0009-0004-0951-2798}
\affiliation{%
	\institution{JD.COM}
	\city{Beijing}
	\country{China}}
\email{chenjianmin.23@jd.com}

\author{Zhuojian Xiao}
\orcid{0009-0000-8875-5771}
\affiliation{%
	\institution{JD.COM}
	\city{Beijing}
	\country{China}}
\email{xiaozhuojian5@jd.com}

\author{Dongyue Wang}
\orcid{0009-0002-4775-6211}
\affiliation{%
	\institution{JD.COM}
	\city{Beijing}
	\country{China}}
\email{wangdongyue@jd.com}

\author{Xiwei Zhao}
\orcid{0000-0002-9382-6041}
\affiliation{%
	\institution{JD.COM}
	\city{Beijing}
	\country{China}}
\email{zhaoxiwei@jd.com}


\renewcommand{\shortauthors}{Zhen-Lin Chen et al.}

\begin{abstract}

Multimodal information is pivotal for e-commerce search ranking. Existing works leverage multimodal data typically by fine-tuning general Multimodal Large Language Models (MLLMs) via collaborative signals, subsequently integrating the derived representations into ranking models as item features. Despite their efficacy, these methods face two primary limitations: (1) they rely on a single collaborative signal for MLLM fine-tuning, failing to exploit the heterogeneous signals essential for multitask ranking; and (2) they treat multimodal representations as regular item features in ranking models, underutilizing their latent potential for user behavior modeling. To address these challenges, we propose the Multiplex Multimodal Representation Model (MMRM), a unified framework that aligns MLLMs with diverse collaborative signals. By employing a shared backbone with task-specific tokens and projection layers, MMRM simultaneously learns from multiple signals and generates comprehensive multiplex item representations in a single inference pass. Furthermore, we introduce a multiplex user representation strategy in ranking models, which derives task-specific user representations via search-based behavior sequence modeling leveraging multiplex item representations. Extensive experiments demonstrate MMRM's superior efficiency and effectiveness. Notably, MMRM has been successfully deployed in the JD e-commerce search engine, yielding significant performance gains for millions of daily users.

\end{abstract}


%

\maketitle

\section{Introduction}

\begin{figure*}
	\centering
	\includegraphics[width=0.92\textwidth]{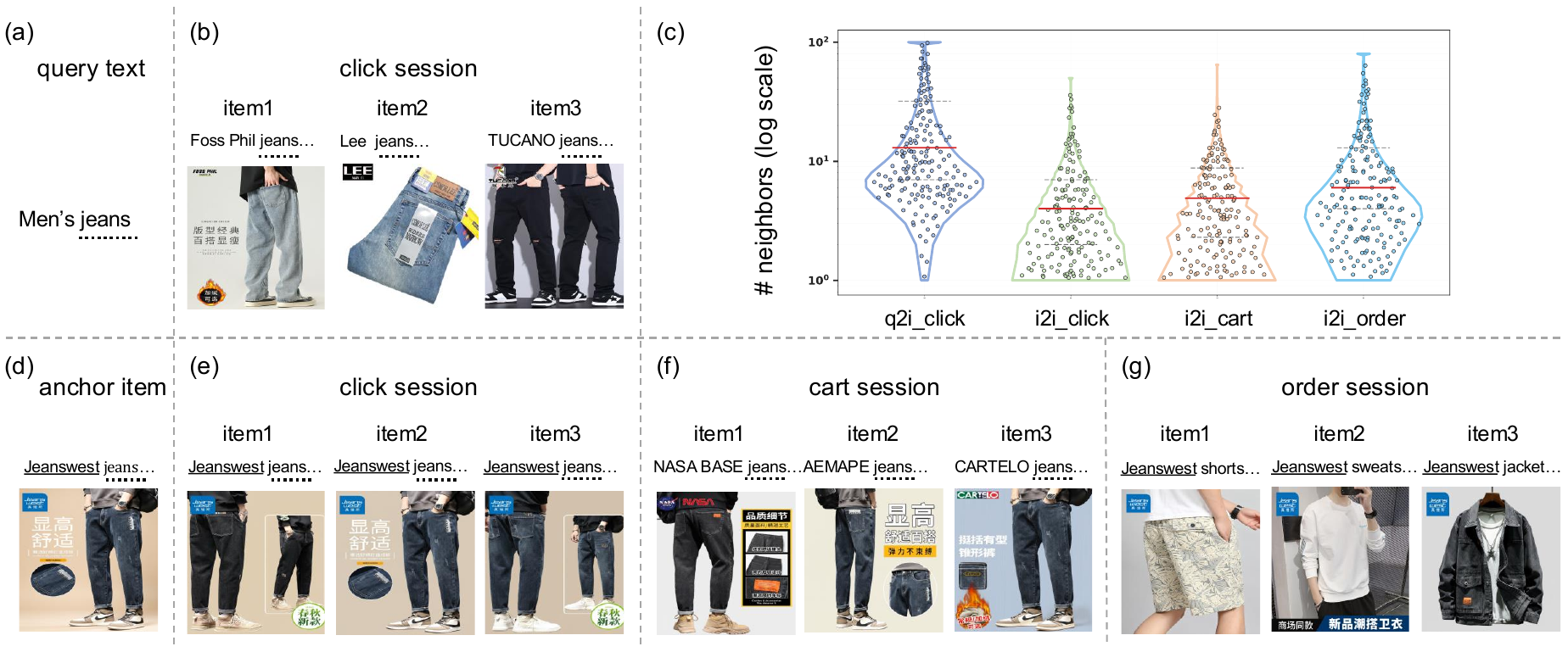}
	\caption{Illustration of heterogeneous collaborative signals. (a-b) Search-based q2i signals; (d-g)  Behavior-based i2i signals within (e) click, (f) cart, and (g) order sessions; (c) Neighbor distribution across different collaborative signals.}
	\label{Fig-Illustration-of-collaborative-signals}
\end{figure*}

E-commerce search engines rely on ranking models to capture fine-grained user preferences across multiple objectives, such as Click-Through Rate (CTR), Add-to-Cart Rate (ACR), and Conversion Rate (CVR). Conventional ID-based ranking models inherently suffer from data sparsity and an inability to perceive product semantics (e.g., titles and images)\cite{fb-dlrm2019,id-vs-mm-sigir23}. To address this, recent works fine-tune general Multimodal Large Language Models (MLLMs) using collaborative signals from e-commerce, then integrate the resulting representations into ranking models as auxiliary features\cite{mm-survey1-kdd24,mm-survey2-arxiv25}.

Despite these advancements, two limitations persist. First, existing methods rely on a single collaborative signal for MLLM fine-tuning, such as search-based query-to-item (q2i) or behavior-based item-to-item (i2i) clicks \cite{taobao-simter-cikm24, kuaishou-qarm-cikm25, xhs-notellm-3w24}. We argue that e-commerce signals are inherently heterogeneous: while q2i clicks capture broad relevance (Fig. \ref{Fig-Illustration-of-collaborative-signals}a-c), sequential i2i behaviors—especially cart and order signals—reflect stronger fine-grained similarities and complementary relationships (Fig. \ref{Fig-Illustration-of-collaborative-signals}d-g). Neglecting these diverse signals prevents MLLMs from learning comprehensive representations essential for multitask ranking. Second, multimodal representations are underutilized in downstream ranking models. Most ranking models introduce one multimodal embedding table for hand-crafted feature interactions \cite{taobao-simter-cikm24,jd-mm-cikm24} or simple behavior grouping \cite{kuaishou-twinv2-cikm24,taobao-sim-mm25}, failing to capture multiplex user intents. This leads to information loss and embedding entanglement in multitask scenarios\cite{tx-multi-emb-kdd24,tx-multi-emb-icml24}.

To address these challenges, we propose the Multiplex Multimodal Representation Model (MMRM), a unified framework aligning MLLMs with four diverse collaborative signals. By employing a shared backbone with task-specific tokens and projection layers, MMRM simultaneously learns from four signals without performance degradation, and efficiently generates four disentangled item representations in a single inference pass. Furthermore, we introduce a multiplex user representation strategy in ranking models, which derives task-specific user representations via search-based behavior sequence modeling using multiplex item representations. Integrating these user representations into corresponding task towers significantly enhances multitask ranking performance.

Our primary contributions are summarized as follows:
\begin{itemize}
\item We propose MMRM, a unified framework that effectively aligns MLLMs with diverse collaborative signals via a shared backbone and task-specific tokens and projection layers, enabling efficient generation of multiplex item representations.

\item We propose a multiplex user representation strategy that derives task-specific user representations from multiplex item representations to boost multitask ranking performance.

\item We demonstrate MMRM's superiority through extensive offline and online experiments, and have deployed it online.
\end{itemize}

\section{Method}

\begin{figure*}
	\centering
	\includegraphics[width=0.96\textwidth]{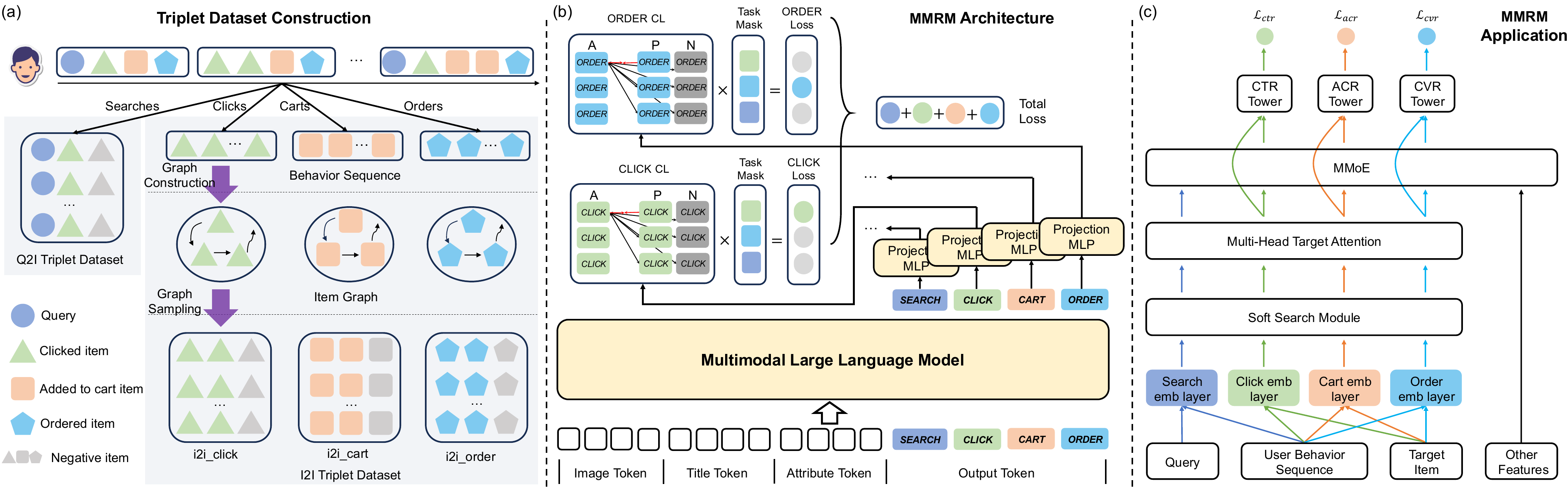}
	\caption{Overview of MMRM. (a) The pipeline for triplet dataset construction; (b) The MMRM architecture; (c) The MMRM application in multitask ranking models.}
	\label{Fig-Overview-of-MMRM}
\end{figure*}

\subsection{Dataset Construction}

\subsubsection{Q2I Dataset}

As shown in Fig. \ref{Fig-Overview-of-MMRM}a, chronological user sessions typically progress from search queries to item clicks, cart additions, and orders, forming a complete behavior sequence. To align textual queries and multimodal items, we construct the q2i\_click dataset using query-item clicks—the most pervasive and coarse-grained collaborative signal. To enhance contrastive learning, we randomly sample a hard negative from the same parent category of the positive but a different sub-category. Our final q2i\_click dataset comprises 0.3 billion diverse triplets collected over six months, capping each query at its top 150 interacted items (Table \ref{tab:dataset}).

\subsubsection{I2I Dataset}

To complement broad q2i signals with fine-grained i2i relationships essential for multitask ranking, we construct three i2i datasets (click, cart, and order) via a graph-based approach (Fig. \ref{Fig-Overview-of-MMRM}a). Using i2i\_click as an example, we construct an item graph by connecting consecutive clicks within a 10-minute window, capturing strong collaborative relations (Table \ref{tab:dataset}). After filtering noisy high-frequency items (promotions, takeaways, and coupons), nodes $v_i$ and edges $e_{ij}$ are weighted by their occurrences $f(v_i)$ and transition frequencies $g(e_{ij})$, respectively. To mitigate hot-item dominance, anchor items $v_a$ are subsampled with probability:

\begin{equation}
	\label{eq-subsampling}
P(v_a)=\sqrt{\frac{t}{f(v_a)}},
\end{equation}
where $t=10^{-5}$ as per word2vec\cite{w2v-nips2013}. For each anchor, we sample up to 30 positive items $v_p$ from its first- and second-order neighbors based on the transition probability:

\begin{equation}
	\label{eq-pos-sampling}
	P(v_p|v_a)=\frac{g(e_{ap})+\sum_{j\in B_a}g(e_{aj})g(e_{jp})}{\sum_{j\in B_a}g(e_{aj})+\sum_{j\in B_a}\sum_{k\in B_j}g(e_{aj})g(e_{jk})},
\end{equation}
where $B_a$ denotes the neighbor set of $v_a$. Finally, we sample hard negatives $v_n$ from the same sub-category as $v_p$, with probability\cite{w2v-nips2013}:

\begin{equation}
	\label{eq-neg-sampling}
	P(v_n|<v_a,v_p>,n\notin\{a,p\})=\frac{f(v_n)^{3/4}}{\sum_{j\in U_{s(v_p)}}f(v_j)^{3/4}},
\end{equation}
where $s(v_p)$ is the sub-category of $v_p$, and $U_{s(v_p)}$ is the item set having the same sub-category as $v_p$.

The i2i\_cart and i2i\_order datasets follow the same generation process of i2i\_click, differing only in configurations (Table \ref{tab:dataset}). Unlike traditional CF-based methods, we capture fine-grained i2i relationships by constructing direct item graphs from sequential user behaviors. Additionally, hard negative sampling further enhances contrastive learning.

\begin{table}
	\caption{Statistics of datasets for MMRM training and testing}
	\label{tab:dataset}
	\begin{tabular}{cccccc}
		\toprule
		\makecell[c]{Data\\ type}&\makecell[c]{Time\\ period}&\makecell[c]{Session\\ window}&\makecell[c]{Max\\ neighbors}&\makecell[c]{\#Train\\Samples}&\makecell[c]{\#Test\\Samples}\\
		\midrule
		q2i\_click & 6 mo & - & 150 & 0.3B & 0.5M\\
		i2i\_click & 6 mo & 10 min & 30 & 0.3B & 0.5M\\
		i2i\_cart & 1 yr & 30 min & 50 & 0.3B & 0.5M\\
		i2i\_order& 3 yr & 7 d & 100 & 0.3B & 0.5M\\
		\bottomrule
	\end{tabular}
\end{table}

\subsection{The Multiplex Multimodal Representation Model}

\subsubsection{Model Architecture}

MMRM takes one q2i and three i2i datasets in a unified triplet format (Fig. \ref{Fig-Overview-of-MMRM}b). Item inputs include an image, a title, and attributes (e.g., shop name, price), while queries are treated as special items without images and attributes. To disentangle tasks, four special tokens—[SEARCH], [CLICK], [CART], and [ORDER]—are appended to the inputs. Leveraging the autoregressive nature of MLLMs, the task-specific embedding for item $i$ is obtained by projecting the last hidden state of special token $t$ via a task-specific MLP:

\begin{equation}
	\label{eq-emb}
	\mathbf{r}_i^t=\text{MLP}^t(\mathbf{h}_i^t), t\in\{[SEARCH], [CLICK], [CART], [ORDER]\}.
\end{equation}

\subsubsection{Multitask Learning Objectives}

During training, four datasets are mixed uniformly. For each task $t$, only its corresponding positive pairs are used for loss calculation, while items from other tasks act as in-batch negatives. The contrastive learning loss for task $t$ is\cite{infonce-loss-19}:

\begin{equation}
	\label{eq-cl-loss-single}
	\mathcal{L}_t^{CL} =-\frac{1}{\sum\limits_{i=1}^B\mathds{1}_t(i)}\sum_{i=1}^B\mathds{1}_t(i)\log\frac{e^{\mathbf{r}_{i^a}^t\cdot\mathbf{r}_{i^p}^t/\tau^t}}{\sum\limits_{j=1}^Be^{\mathbf{r}_{i^a}^t\cdot\mathbf{r}_{j^p}^t/\tau^t}+\sum\limits_{j=1}^Be^{\mathbf{r}_{i^a}^t\cdot\mathbf{r}_{j^n}^t/\tau^t}},
\end{equation}
where $B$ is the batch size; $\mathds{1}_t(i)$ is an indicator function returning 1 if the $i$-th instance belongs to task $t$; $\mathbf{r}_{i^a}^t$, $\mathbf{r}_{i^p}^t$, and $\mathbf{r}_{i^n}^t$ are anchor, positive, and negative embeddings, respectively, for the $i$-th instance from task $t$; and $\tau^t$ is the temperature for task $t$. Finally, the total loss aggregates these task-specific losses with weights $\gamma^t$:

\begin{equation}
	\label{eq-cl-loss-total}
	\mathcal{L}_{total}^{CL} =\sum\limits_{t}\gamma^t\mathcal{L}_t^{CL}, t\in\{[SEARCH], [CLICK], [CART], [ORDER]\}.
\end{equation}

\subsection{The Multitask Ranking Model}

As illustrated in Fig. \ref{Fig-Overview-of-MMRM}c, given a query $Q$, a user behavior sequence $S$, a target item $I$, and other features $O$, we first extract a top-K relevant behavior subsequence $S'_t$ using the multimodal embedding table specific to task $t$:

\begin{equation}
	\label{eq-rank-soft-search-i}
	S_{t}'=\text{Soft-Search}(S, I/Q, K, t).
\end{equation}

Then, we derive a task-specific user representation $U_t$ using a Multi-Head Target Attention (MHTA) module (multi-head DIN\cite{taobao-din-kdd18}):

\begin{equation}
	\label{eq-rank-mhta-i}
	U_{t}=\text{MHTA}^t(S_{t}', I/Q, t).
\end{equation}

After that, all user representations and features $O$ are processed by an MMoE module\cite{mmoe-kdd18} to yield the task-specific hidden state $X_t$:

\begin{equation}
	\label{eq-rank-mmoe}
	X_{t}=\text{MMoE}(U_{[SEARCH]}, U_{[CLICK]}, U_{[CART]}, U_{[ORDER]}, O, t).
\end{equation}

At last, to predict for task $t$, $X_{t}$ and the task-specific user representation $U_{t}$ are jointly fed into an MLP-based tower network:

\begin{equation}
	\label{eq-rank-output}
	\hat{y}_{t}=\text{Tower}^t(X_{t}, U_{t}).
\end{equation}

The overall multitask ranking loss is defined as a weighted sum of task-specific cross-entropy losses\cite{ce-loss-icml23}:

\begin{equation}
	\label{eq-rank-loss-single}
	\mathcal{L}_{t}^{rank}=\text{Cross-entropy}(y_t,\hat{y}_t),
\end{equation}

\begin{equation}
	\label{eq-rank-loss-total}
	\mathcal{L}_{total}^{rank} =\sum\limits_{t}\lambda^t\mathcal{L}_t^{rank}, t\in\{ctr, acr, cvr\},
\end{equation}
where $\lambda^t$ is the loss weight for task $t$.

\section{Experiments}

\subsection{Experimental Setup}

\subsubsection{Datasets and Metrics}

\begin{table*}[t]
	\caption{Evaluation results for representation models. Bold and underline denote the best and second-best results, respectively.}
	\label{tab:pretrain-main-results}
	\begin{tabular}{cc|cccc|cccc}
		\hline
		\multicolumn{2}{c|}{\multirow{2}{*}{\diagbox[width=10em]{Model}{Task}}} & \multicolumn{4}{c|}{F1@5}                        & \multicolumn{4}{c}{NDCG@5}                       \\ \cline{3-10} 
		\multicolumn{2}{c|}{}                                         & q2i\_click & i2i\_click & i2i\_cart & i2i\_order & q2i\_click & i2i\_click & i2i\_cart & i2i\_order \\ \hline
		\multirow{4}{*}{Single}              & q2i\_click             & \underline{0.1985}     & 0.2943     & 0.2679    & 0.0675     &  \underline{0.6356}     & 0.7289     & 0.6655    & 0.4752     \\
		& i2i\_click             & 0.0818     & \underline{0.3254}     & 0.2981    & 0.0831     & 0.6129     & \underline{0.7367}     & 0.6752    & 0.4976     \\
		& i2i\_cart              & 0.0238     & 0.3237     & \underline{0.3013}    & \underline{0.0908}     & 0.5948     & 0.7351     & \underline{0.6763}    & 0.5070     \\
		& i2i\_order             & 0.0087     & 0.2989     & 0.2809    & 0.0904     & 0.5579     & 0.7293     & 0.6710    & \underline{0.5224}     \\ \hline
		\multicolumn{2}{c|}{Vanilla-Multi}                            & 0.1593     & 0.2987     & 0.2748    & 0.0676     & 0.6127     & 0.7295     & 0.6702    & 0.4875     \\ \hline
		\multicolumn{2}{c|}{MMRM}                                     & \textbf{0.2055}     & \textbf{0.3276}     & \textbf{0.3037}    & \textbf{0.0934}     & \textbf{0.6392}     & \textbf{0.7371}     & \textbf{0.6779}    & \textbf{0.5249}     \\ \hline
	\end{tabular}
\end{table*}

We evaluated MMRM's multitask performance on four held-out test sets (0.5M samples each, Table \ref{tab:dataset}). After models were trained, we derived embeddings used in test sets, performed ANN search tasks, and measured retrieval and ranking performance via F1@k\cite{f12015} and NDCG@k\cite{ndcg2002}, respectively. To evaluate ranking models, we collected 8-day JD search logs, and used the first 7 days for training (4B samples) and the last day for testing (2M samples). Following established practices, ranking performance was assessed via GAUC\cite{taobao-gauc-kdd17,taobao-din-kdd18}.

\subsubsection{Baseline Models}

We compared three types of representation models: \textbf{Single} (task-specific models), \textbf{Vanilla-Multi} (multitask via task-specific prompts and a shared [EMB] token), and \textbf{MMRM} (multitask via task-specific tokens). Besides, we compared four types of SIM-based ranking models\cite{taobao-sim-cikm20}: $\boldsymbol{\text{\textbf{SIM}}_{hard}(category)}$ (category-based behavior filtering), $\boldsymbol{\text{\textbf{SIM}}_{soft}(item_{e2e})}$ (filtering by end-to-end item embeddings; our online baseline\cite{jd-awmoe-icde23}), $\boldsymbol{\text{\textbf{SIM}}_{soft}(item_{MMRM[t]})}$ (filtering by MMRM's task $t$ embeddings), and $\boldsymbol{\text{\textbf{SIM}}_{soft}(item_{MMRM}}$\linebreak$\boldsymbol{{}_{[ALL]})}$ (multiplex filtering by all embeddings from MMRM).

\subsubsection{Implementation Details}

MMRM initialized from Qwen3-VL-4B-Instruct\cite{bai2025qwen3vltechnicalreport}, training all parameters except the vision encoder. Training used Adam\cite{adam-iclr2015} (learning rate 2e-5, batch size 2048 via GradCache\cite{gao-etal-2021-scaling}), FlashAttention-2\cite{dao2023flashattention2fasterattentionbetter}, $\tau^t=0.05$, and $\gamma^t=0.25$. Ranking models used AdamW\cite{loshchilov2018decoupled} (learning rate 3e-5, batch size 4096) and $\lambda^t=0.33$. All models were trained on 8 NVIDIA H800 GPUs.

\subsection{Evaluations for Representation Models}

\subsubsection{Main Results}

Evaluation results for representation models (Table \ref{tab:pretrain-main-results}) reveal three insights: (1) Single task models excelled only on their specific training tasks, highlighting task heterogeneity and the inability of a universal embedding to align all collaborative signals; (2) Vanilla-Multi underperformed all single-task baselines, indicating that task-specific prompts with a shared [EMB] token fail to disentangle tasks; (3) MMRM consistently outperformed all baselines, validating its architecture design. MMRM's superiority over single task models likely stems from more diverse in-batch negatives during multitask training. Additionally, distinct task tokens enable MMRM to efficiently generate all four task embeddings in one forward pass, avoiding four separate passes required by baselines.

\subsubsection{Ablation Study}

Contrastive learning typically benefits from a large batch size\cite{cl-large-bs-nips22}, which is often bottlenecked by the high GPU memory consumption of modern MLLMs. To scale the batch size, we employed GradCache\cite{gao-etal-2021-scaling}, which splits the backpropagation of the contrastive loss into two stages and breaks the model update into memory-efficient sub-updates. This enabled us to train the 4B-parameter MMRM with a batch size of 4096 on a single H800 machine. Our ablation study on the batch size (Table \ref{tab:ablation-bs-results}) confirms that larger batch sizes consistently boost MMRM's performance.

\begin{table}[]
	\caption{Ablation study on batch size. GradCache was used for batch sizes > 1024, and images were omitted for acceleration.}
	\label{tab:ablation-bs-results}
	\begin{tabular}{c|cccc}
		\hline
		\multirow{2}{*}{\diagbox[width=8em]{Batch size}{Task}} & \multicolumn{4}{c}{F1@5}                         \\ \cline{2-5} 
		& q2i\_click & i2i\_click & i2i\_cart & i2i\_order \\ \hline
		256                                            & 0.1874     & 0.3216     & 0.2991    & 0.0882     \\
		512                                            & 0.1937     & 0.3223     & 0.2984    & 0.0895     \\
		1024                                           & 0.1980     & 0.3239     & 0.2991    & 0.0907     \\
		2048      & \textbf{0.2007}     & \textbf{0.3255}     & 0.3004    & 0.0916     \\
		4096      & 0.1997     & 0.3251     & \textbf{0.3021}    & \textbf{0.0919}     \\ \hline
	\end{tabular}
\end{table}

\subsection{Evaluations for Ranking Models}

Table \ref{tab:ranking-main-results} presents evaluation results for ranking models. Our online baseline, $\text{SIM}_{soft}(item_{e2e})$, outperformed $\text{SIM}_{hard}(category)$, validating the soft search strategy. However, $\text{SIM}_{soft}(item_{e2e})$ relies on a single end-to-end embedding table for behavior filtering and user modeling. This risks embedding entanglement, thereby limiting performance in multitask scenarios. By incorporating an additional multimodal embedding table for soft search, $\text{SIM}_{soft}(item_{MMRM[t]})$ achieved significantly better results across all three tasks. Ultimately, utilizing all four multimodal embedding tables for task-specific user modeling, our proposed $\text{SIM}_{soft}(item_{MMRM[ALL]})$ attained the best performance, demonstrating the effectiveness of multiplex multimodal representations in multitask ranking.

\begin{table}[]
	\caption{Evaluation results for ranking models}
	\label{tab:ranking-main-results}
	\begin{tabular}{c|ccc}
		\hline
		\multirow{2}{*}{\diagbox[width=13em]{Model}{Task}} & \multicolumn{3}{c}{GAUC} \\ \cline{2-4} 
		& CTR    & ACR    & CVR    \\ \hline
		$\text{SIM}_{hard}(category)$                              & 0.6816 & 0.7051 & 0.6733 \\ \hline
		$\text{SIM}_{soft}(item_{e2e})$                              & 0.6866 & 0.7101 & 0.6783 \\ \hline
		$\text{SIM}_{soft}(item_{MMRM[SEARCH]})$                      & 0.6947 & 0.7160 & 0.6807 \\
		$\text{SIM}_{soft}(item_{MMRM[CLICK]})$                       & 0.6928 & 0.7145 & 0.6839 \\
		$\text{SIM}_{soft}(item_{MMRM[CART]})$                        & 0.6941 & 0.7124 & 0.6830 \\
		$\text{SIM}_{soft}(item_{MMRM[ORDER]})$                      & 0.6934 & 0.7127 & 0.6814 \\ \hline
		$\text{SIM}_{soft}(item_{MMRM[ALL]})$                         & \textbf{0.7037}   & \textbf{0.7190}   & \textbf{0.6875} \\ \hline
	\end{tabular}
\end{table}

\subsection{Online A/B Test}

In a week-long online A/B test on the JD e-commerce search engine, our proposed  $\text{SIM}_{soft}(item_{MMRM[ALL]})$ outperformed the online baseline $\text{SIM}_{soft}(item_{e2e})$, improving user click-through (UCTR), add-to-cart (UACR), and conversion (UCVR) rates by 0.42\%, 0.37\%, and 0.35\%, respectively. Given the massive traffic of the platform, even a 0.10\% lift in these metrics yields substantial revenue growth, making these gains highly significant. Consequently, both MMRM and the proposed ranking model have been fully deployed online.

\section{Conclusions}

In this paper, we proposed the Multiplex Multimodal Representation Model (MMRM) to integrate multimodal information into ranking models. By coupling a shared backbone with task-specific tokens and projection layers, MMRM simultaneously aligns with four graph-derived collaborative signals without performance degradation, generating comprehensive multiplex item representations in a single inference pass. Furthermore, we introduced a multiplex user representation strategy leveraging multiplex item representations, boosting downstream multitask ranking performance. Extensive offline and online experiments validated the effectiveness of MMRM and the proposed ranking model.

\section{Presenter Biography}
\textbf{Zhen-Lin Chen} is an applied scientist at JD.COM, focusing on research and development of large-scale search and recommendation systems to improve user online shopping experiences. He received his doctoral degree from the Institute of Computing Technology, Chinese Academy of Sciences. His research mainly focuses on natural language processing and information retrieval.

\bibliographystyle{ACM-Reference-Format}
\bibliography{sample-base}

@STRING{may = "May"}

@STRING{jun = "June"}

@STRING{aug = "Aug."}

@STRING{oct = "Oct."}

@misc{fb-dlrm2019,
	title={Deep Learning Recommendation Model for Personalization and Recommendation Systems}, 
	author={Maxim Naumov and Dheevatsa Mudigere and Hao-Jun Michael Shi and Jianyu Huang and Narayanan Sundaraman and Jongsoo Park and Xiaodong Wang and Udit Gupta and Carole-Jean Wu and Alisson G. Azzolini and Dmytro Dzhulgakov and Andrey Mallevich and Ilia Cherniavskii and Yinghai Lu and Raghuraman Krishnamoorthi and Ansha Yu and Volodymyr Kondratenko and Stephanie Pereira and Xianjie Chen and Wenlin Chen and Vijay Rao and Bill Jia and Liang Xiong and Misha Smelyanskiy},
	year={2019},
	eprint={1906.00091},
	archivePrefix={arXiv},
	primaryClass={cs.IR},
	url={https://arxiv.org/abs/1906.00091}, 
}

@inproceedings{id-vs-mm-sigir23,
	author = {Yuan, Zheng and Yuan, Fajie and Song, Yu and Li, Youhua and Fu, Junchen and Yang, Fei and Pan, Yunzhu and Ni, Yongxin},
	title = {Where to Go Next for Recommender Systems? ID- vs. Modality-based Recommender Models Revisited},
	year = {2023},
	isbn = {9781450394086},
	publisher = {Association for Computing Machinery},
	address = {New York, NY, USA},
	url = {https://doi.org/10.1145/3539618.3591932},
	doi = {10.1145/3539618.3591932},
	booktitle = {Proceedings of the 46th International ACM SIGIR Conference on Research and Development in Information Retrieval},
	pages = {2639–2649},
	numpages = {11},
	location = {Taipei, Taiwan},
	series = {SIGIR '23}
}

@inproceedings{mm-survey1-kdd24,
	author = {Liu, Qijiong and Zhu, Jieming and Yang, Yanting and Dai, Quanyu and Du, Zhaocheng and Wu, Xiao-Ming and Zhao, Zhou and Zhang, Rui and Dong, Zhenhua},
	title = {Multimodal Pretraining, Adaptation, and Generation for Recommendation: A Survey},
	year = {2024},
	isbn = {9798400704901},
	publisher = {Association for Computing Machinery},
	address = {New York, NY, USA},
	url = {https://doi.org/10.1145/3637528.3671473},
	doi = {10.1145/3637528.3671473},
	booktitle = {Proceedings of the 30th ACM SIGKDD Conference on Knowledge Discovery and Data Mining},
	pages = {6566–6576},
	numpages = {11},
	location = {Barcelona, Spain},
	series = {KDD '24}
}

@misc{mm-survey2-arxiv25,
	title={A Survey on Large Language Models in Multimodal Recommender Systems}, 
	author={Alejo Lopez-Avila and Jinhua Du},
	year={2025},
	eprint={2505.09777},
	archivePrefix={arXiv},
	primaryClass={cs.IR},
	url={https://arxiv.org/abs/2505.09777}, 
}

@inproceedings{jd-mm-cikm24,
	author = {Xu, Enqiang and Li, Xinhui and Zhou, Zhigong and Ji, Jiahao and Zhao, Jinyuan and Miao, Dadong and Wang, Songlin and Liu, Lin and Xu, Sulong},
	title = {Advancing Re-Ranking with Multimodal Fusion and Target-Oriented Auxiliary Tasks in E-Commerce Search},
	year = {2024},
	isbn = {9798400704369},
	publisher = {Association for Computing Machinery},
	address = {New York, NY, USA},
	url = {https://doi.org/10.1145/3627673.3680063},
	doi = {10.1145/3627673.3680063},
	booktitle = {Proceedings of the 33rd ACM International Conference on Information and Knowledge Management},
	pages = {5007–5014},
	numpages = {8},
	location = {Boise, ID, USA},
	series = {CIKM '24}
}

@inproceedings{taobao-simter-cikm24,
	author = {Sheng, Xiang-Rong and Yang, Feifan and Gong, Litong and Wang, Biao and Chan, Zhangming and Zhang, Yujing and Cheng, Yueyao and Zhu, Yong-Nan and Ge, Tiezheng and Zhu, Han and Jiang, Yuning and Xu, Jian and Zheng, Bo},
	title = {Enhancing Taobao Display Advertising with Multimodal Representations: Challenges, Approaches and Insights},
	year = {2024},
	isbn = {9798400704369},
	publisher = {Association for Computing Machinery},
	address = {New York, NY, USA},
	url = {https://doi.org/10.1145/3627673.3680068},
	doi = {10.1145/3627673.3680068},
	booktitle = {Proceedings of the 33rd ACM International Conference on Information and Knowledge Management},
	pages = {4858–4865},
	numpages = {8},
	location = {Boise, ID, USA},
	series = {CIKM '24}
}

@inproceedings{kuaishou-qarm-cikm25,
	author = {Luo, Xinchen and Cao, Jiangxia and Sun, Tianyu and Yu, Jinkai and Huang, Rui and Yuan, Wei and Lin, Hezheng and Zheng, Yichen and Wang, Shiyao and Hu, Qigen and Qiu, Changqing and Zhang, Jiaqi and Zhang, Xu and Yan, Zhiheng and Zhang, Jingming and Zhang, Simin and Wen, Mingxing and Liu, Zhaojie and Zhou, Guorui},
	title = {QARM: Quantitative Alignment Multi-Modal Recommendation at Kuaishou},
	year = {2025},
	isbn = {9798400720406},
	publisher = {Association for Computing Machinery},
	address = {New York, NY, USA},
	url = {https://doi.org/10.1145/3746252.3761502},
	doi = {10.1145/3746252.3761502},
	booktitle = {Proceedings of the 34th ACM International Conference on Information and Knowledge Management},
	pages = {5915–5922},
	numpages = {8},
	location = {Seoul, Republic of Korea},
	series = {CIKM '25}
}

@misc{taobao-sim-mm25,
	title={MUSE: A Simple Yet Effective Multimodal Search-Based Framework for Lifelong User Interest Modeling}, 
	author={Bin Wu and Feifan Yang and Zhangming Chan and Yu-Ran Gu and Jiawei Feng and Chao Yi and Xiang-Rong Sheng and Han Zhu and Jian Xu and Mang Ye and Bo Zheng},
	year={2025},
	eprint={2512.07216},
	archivePrefix={arXiv},
	primaryClass={cs.IR},
	url={https://arxiv.org/abs/2512.07216}, 
}

@inproceedings{xhs-notellm-3w24,
	author = {Zhang, Chao and Wu, Shiwei and Zhang, Haoxin and Xu, Tong and Gao, Yan and Hu, Yao and Chen, Enhong},
	title = {NoteLLM: A Retrievable Large Language Model for Note Recommendation},
	year = {2024},
	isbn = {9798400701726},
	publisher = {Association for Computing Machinery},
	address = {New York, NY, USA},
	url = {https://doi.org/10.1145/3589335.3648314},
	doi = {10.1145/3589335.3648314},
	booktitle = {Companion Proceedings of the ACM Web Conference 2024},
	pages = {170–179},
	numpages = {10},
	location = {Singapore, Singapore},
	series = {WWW '24}
}

@inproceedings{kuaishou-twinv2-cikm24,
	author = {Si, Zihua and Guan, Lin and Sun, Zhongxiang and Zang, Xiaoxue and Lu, Jing and Hui, Yiqun and Cao, Xingchao and Yang, Zeyu and Zheng, Yichen and Leng, Dewei and Zheng, Kai and Zhang, Chenbin and Niu, Yanan and Song, Yang and Gai, Kun},
	title = {TWIN V2: Scaling Ultra-Long User Behavior Sequence Modeling for Enhanced CTR Prediction at Kuaishou},
	year = {2024},
	isbn = {9798400704369},
	publisher = {Association for Computing Machinery},
	address = {New York, NY, USA},
	url = {https://doi.org/10.1145/3627673.3680030},
	doi = {10.1145/3627673.3680030},
	booktitle = {Proceedings of the 33rd ACM International Conference on Information and Knowledge Management},
	pages = {4890–4897},
	numpages = {8},
	location = {Boise, ID, USA},
	series = {CIKM '24}
}

@inproceedings{tx-multi-emb-kdd24,
	author = {Pan, Junwei and Xue, Wei and Wang, Ximei and Yu, Haibin and Liu, Xun and Quan, Shijie and Qiu, Xueming and Liu, Dapeng and Xiao, Lei and Jiang, Jie},
	title = {Ads Recommendation in a Collapsed and Entangled World},
	year = {2024},
	isbn = {9798400704901},
	publisher = {Association for Computing Machinery},
	address = {New York, NY, USA},
	url = {https://doi.org/10.1145/3637528.3671607},
	doi = {10.1145/3637528.3671607},
	booktitle = {Proceedings of the 30th ACM SIGKDD Conference on Knowledge Discovery and Data Mining},
	pages = {5566–5577},
	numpages = {12},
	location = {Barcelona, Spain},
	series = {KDD '24}
}

@inproceedings{tx-multi-emb-icml24,
	author = {Guo, Xingzhuo and Pan, Junwei and Wang, Ximei and Chen, Baixu and Jiang, Jie and Long, Mingsheng},
	title = {On the embedding collapse when scaling up recommendation models},
	year = {2024},
	publisher = {JMLR.org},
	booktitle = {Proceedings of the 41st International Conference on Machine Learning},
	articleno = {671},
	numpages = {19},
	location = {Vienna, Austria},
	series = {ICML'24}
}

@misc{infonce-loss-19,
	title={Representation Learning with Contrastive Predictive Coding}, 
	author={Aaron van den Oord and Yazhe Li and Oriol Vinyals},
	year={2019},
	eprint={1807.03748},
	archivePrefix={arXiv},
	primaryClass={cs.LG},
	url={https://arxiv.org/abs/1807.03748}, 
}

@inproceedings{ce-loss-icml23,
	author = {Mao, Anqi and Mohri, Mehryar and Zhong, Yutao},
	title = {Cross-entropy loss functions: theoretical analysis and applications},
	year = {2023},
	publisher = {JMLR.org},
	booktitle = {Proceedings of the 40th International Conference on Machine Learning},
	articleno = {992},
	numpages = {26},
	location = {Honolulu, Hawaii, USA},
	series = {ICML'23}
}

@inproceedings{w2v-nips2013,
	author = {Mikolov, Tomas and Sutskever, Ilya and Chen, Kai and Corrado, Greg S and Dean, Jeff},
	booktitle = {Advances in Neural Information Processing Systems},
	editor = {C.J. Burges and L. Bottou and M. Welling and Z. Ghahramani and K.Q. Weinberger},
	pages = {},
	publisher = {Curran Associates, Inc.},
	title = {Distributed Representations of Words and Phrases and their Compositionality},
	url = {https://proceedings.neurips.cc/paper_files/paper/2013/file/9aa42b31882ec039965f3c4923ce901b-Paper.pdf},
	volume = {26},
	year = {2013}
}

@inproceedings{taobao-din-kdd18,
	author = {Zhou, Guorui and Zhu, Xiaoqiang and Song, Chenru and Fan, Ying and Zhu, Han and Ma, Xiao and Yan, Yanghui and Jin, Junqi and Li, Han and Gai, Kun},
	title = {Deep Interest Network for Click-Through Rate Prediction},
	year = {2018},
	isbn = {9781450355520},
	publisher = {Association for Computing Machinery},
	address = {New York, NY, USA},
	url = {https://doi.org/10.1145/3219819.3219823},
	doi = {10.1145/3219819.3219823},
	booktitle = {Proceedings of the 24th ACM SIGKDD International Conference on Knowledge Discovery \& Data Mining},
	pages = {1059–1068},
	numpages = {10},
	location = {London, United Kingdom},
	series = {KDD '18}
}

@inproceedings{mmoe-kdd18,
	author = {Ma, Jiaqi and Zhao, Zhe and Yi, Xinyang and Chen, Jilin and Hong, Lichan and Chi, Ed H.},
	title = {Modeling Task Relationships in Multi-task Learning with Multi-gate Mixture-of-Experts},
	year = {2018},
	isbn = {9781450355520},
	publisher = {Association for Computing Machinery},
	address = {New York, NY, USA},
	url = {https://doi.org/10.1145/3219819.3220007},
	doi = {10.1145/3219819.3220007},
	booktitle = {Proceedings of the 24th ACM SIGKDD International Conference on Knowledge Discovery \& Data Mining},
	pages = {1930–1939},
	numpages = {10},
	location = {London, United Kingdom},
	series = {KDD '18}
}

@ARTICLE{f12015,
	author={Huang, Hao and Xu, Haihua and Wang, Xianhui and Silamu, Wushour},
	journal={IEEE/ACM Transactions on Audio, Speech, and Language Processing}, 
	title={Maximum F1-Score Discriminative Training Criterion for Automatic Mispronunciation Detection}, 
	year={2015},
	volume={23},
	number={4},
	pages={787-797},
	doi={10.1109/TASLP.2015.2409733}}

@article{ndcg2002,
	author = {J\"{a}rvelin, Kalervo and Kek\"{a}l\"{a}inen, Jaana},
	title = {Cumulated gain-based evaluation of IR techniques},
	year = {2002},
	issue_date = {October 2002},
	publisher = {Association for Computing Machinery},
	address = {New York, NY, USA},
	volume = {20},
	number = {4},
	issn = {1046-8188},
	url = {https://doi.org/10.1145/582415.582418},
	doi = {10.1145/582415.582418},
	journal = {ACM Trans. Inf. Syst.},
	month = oct,
	pages = {422–446},
	numpages = {25}
}

@inproceedings{taobao-gauc-kdd17,
	author = {Zhu, Han and Jin, Junqi and Tan, Chang and Pan, Fei and Zeng, Yifan and Li, Han and Gai, Kun},
	title = {Optimized Cost per Click in Taobao Display Advertising},
	year = {2017},
	isbn = {9781450348874},
	publisher = {Association for Computing Machinery},
	address = {New York, NY, USA},
	url = {https://doi.org/10.1145/3097983.3098134},
	doi = {10.1145/3097983.3098134},
	booktitle = {Proceedings of the 23rd ACM SIGKDD International Conference on Knowledge Discovery and Data Mining},
	pages = {2191–2200},
	numpages = {10},
	location = {Halifax, NS, Canada},
	series = {KDD '17}
}

@inproceedings{taobao-sim-cikm20,
	author = {Pi, Qi and Zhou, Guorui and Zhang, Yujing and Wang, Zhe and Ren, Lejian and Fan, Ying and Zhu, Xiaoqiang and Gai, Kun},
	title = {Search-based User Interest Modeling with Lifelong Sequential Behavior Data for Click-Through Rate Prediction},
	year = {2020},
	isbn = {9781450368599},
	publisher = {Association for Computing Machinery},
	address = {New York, NY, USA},
	url = {https://doi.org/10.1145/3340531.3412744},
	doi = {10.1145/3340531.3412744},
	booktitle = {Proceedings of the 29th ACM International Conference on Information \& Knowledge Management},
	pages = {2685–2692},
	numpages = {8},
	location = {Virtual Event, Ireland},
	series = {CIKM '20}
}

@misc{bai2025qwen3vltechnicalreport,
	title={Qwen3-VL Technical Report}, 
	author={Shuai Bai and Yuxuan Cai and Ruizhe Chen and Keqin Chen and Xionghui Chen and Zesen Cheng and Lianghao Deng and Wei Ding and Chang Gao and Chunjiang Ge and Wenbin Ge and Zhifang Guo and Qidong Huang and Jie Huang and Fei Huang and Binyuan Hui and Shutong Jiang and Zhaohai Li and Mingsheng Li and Mei Li and Kaixin Li and Zicheng Lin and Junyang Lin and Xuejing Liu and Jiawei Liu and Chenglong Liu and Yang Liu and Dayiheng Liu and Shixuan Liu and Dunjie Lu and Ruilin Luo and Chenxu Lv and Rui Men and Lingchen Meng and Xuancheng Ren and Xingzhang Ren and Sibo Song and Yuchong Sun and Jun Tang and Jianhong Tu and Jianqiang Wan and Peng Wang and Pengfei Wang and Qiuyue Wang and Yuxuan Wang and Tianbao Xie and Yiheng Xu and Haiyang Xu and Jin Xu and Zhibo Yang and Mingkun Yang and Jianxin Yang and An Yang and Bowen Yu and Fei Zhang and Hang Zhang and Xi Zhang and Bo Zheng and Humen Zhong and Jingren Zhou and Fan Zhou and Jing Zhou and Yuanzhi Zhu and Ke Zhu},
	year={2025},
	eprint={2511.21631},
	archivePrefix={arXiv},
	primaryClass={cs.CV},
	url={https://arxiv.org/abs/2511.21631}, 
}

@misc{dao2023flashattention2fasterattentionbetter,
	title={FlashAttention-2: Faster Attention with Better Parallelism and Work Partitioning}, 
	author={Tri Dao},
	year={2023},
	eprint={2307.08691},
	archivePrefix={arXiv},
	primaryClass={cs.LG},
	url={https://arxiv.org/abs/2307.08691}, 
}

@inproceedings{
	loshchilov2018decoupled,
	title={Decoupled Weight Decay Regularization},
	author={Ilya Loshchilov and Frank Hutter},
	booktitle={International Conference on Learning Representations},
	year={2019},
	url={https://openreview.net/forum?id=Bkg6RiCqY7},
}

@inproceedings{adam-iclr2015,
	author       = {Diederik P. Kingma and
	Jimmy Ba},
	editor       = {Yoshua Bengio and
	Yann LeCun},
	title        = {Adam: {A} Method for Stochastic Optimization},
	booktitle    = {3rd International Conference on Learning Representations, {ICLR} 2015,
	San Diego, CA, USA, May 7-9, 2015, Conference Track Proceedings},
	year         = {2015},
	url          = {http://arxiv.org/abs/1412.6980},
	bibsource    = {dblp computer science bibliography, https://dblp.org}
}

@inproceedings{gao-etal-2021-scaling,
	title = "Scaling Deep Contrastive Learning Batch Size under Memory Limited Setup",
	author = "Gao, Luyu  and
	Zhang, Yunyi  and
	Han, Jiawei  and
	Callan, Jamie",
	editor = "Rogers, Anna  and
	Calixto, Iacer  and
	Vuli{\'c}, Ivan  and
	Saphra, Naomi  and
	Kassner, Nora  and
	Camburu, Oana-Maria  and
	Bansal, Trapit  and
	Shwartz, Vered",
	booktitle = "Proceedings of the 6th Workshop on Representation Learning for NLP (RepL4NLP-2021)",
	month = aug,
	year = "2021",
	address = "Online",
	publisher = "Association for Computational Linguistics",
	url = "https://aclanthology.org/2021.repl4nlp-1.31/",
	doi = "10.18653/v1/2021.repl4nlp-1.31",
	pages = "316--321"
}

@inproceedings{jd-awmoe-icde23,
	author       = {Juan Gong and
	Zhenlin Chen and
	Chaoyi Ma and
	Zhuojian Xiao and
	Haonan Wang and
	Guoyu Tang and
	Lin Liu and
	Sulong Xu and
	Bo Long and
	Yunjiang Jiang},
	title        = {Attention Weighted Mixture of Experts with Contrastive Learning for
	Personalized Ranking in E-commerce},
	booktitle    = {39th {IEEE} International Conference on Data Engineering, {ICDE} 2023,
	Anaheim, CA, USA, April 3-7, 2023},
	pages        = {3222--3234},
	publisher    = {{IEEE}},
	year         = {2023},
	url          = {https://doi.org/10.1109/ICDE55515.2023.00247},
	doi          = {10.1109/ICDE55515.2023.00247},
	bibsource    = {dblp computer science bibliography, https://dblp.org}
}

@inproceedings{cl-large-bs-nips22,
	author = {Chen, Changyou and Zhang, Jianyi and Xu, Yi and Chen, Liqun and Duan, Jiali and Chen, Yiran and Tran, Son Dinh and Zeng, Belinda and Chilimbi, Trishul},
	title = {Why do we need large batchsizes in contrastive learning? a gradient-bias perspective},
	year = {2022},
	isbn = {9781713871088},
	publisher = {Curran Associates Inc.},
	address = {Red Hook, NY, USA},
	booktitle = {Proceedings of the 36th International Conference on Neural Information Processing Systems},
	articleno = {2454},
	numpages = {16},
	location = {New Orleans, LA, USA},
	series = {NIPS '22}
}

%
%
%
%
%
%
%
%

\end{document}